\documentclass[11pt]{article}

\def\be{\begin{equation}}
\def\ee{\end{equation}}
\def\ba{\begin{eqnarray}}
\def\ea{\end{eqnarray}}
\def\la{\langle}
\def\ra{\rangle}

\def\h{\hskip 1cm}

\def\lo{\longrightarrow}

\usepackage{graphicx}
\begin{document}
\begin{titlepage}
\vspace{4cm}
\begin{center}{\Large \bf Entanglement of bosonic modes in symmetric graphs}\\
\vspace{1cm}M. Asoudeh\footnote{email:asoudeh@mehr.sharif.edu},
\hspace{0.5cm} V. Karimipour \footnote{Corresponding author, email:vahid@sharif.edu}\\
\vspace{1cm} Department of Physics, Sharif University of Technology,\\
P.O. Box 11365-9161,\\ Tehran, Iran
\end{center}
\vskip 3cm
\begin{abstract}
The ground and thermal states of a quadratic hamiltonian
representing the interaction of bosonic modes or particles are
always Gaussian states. We investigate the entanglement properties
of these states for the case where the interactions are
represented by harmonic forces acting along the edges of symmetric
graphs, i.e. 1, 2, and 3 dimensional rectangular lattices, mean
field clusters and platonic solids. We determine the Entanglement
of Formation (EoF) as a function of the interaction strength,
calculate the maximum EoF in each case and compare these values
with the bounds found in \cite{wolf} which are valid for any
quadratic hamiltonian.
\end{abstract}
\end{titlepage}

\vskip 3cm
\section{Introduction}\label{intro}
Suppose we have a many body quantum system in its ground state. We
ask how much  quantum correlation or entanglement exists between
two of the particles? How this entanglement depends on the
strength of interaction between the particles? Does it extend far
beyond the nearest neighbor particles? If we raise the
temperature, hence mixing the ground state with higher level
states, at what temperature the entanglement ceases to exist? Some
of these questions have been investigated in recent years for spin
systems \cite{abv, zan, ost1, cw, jin, osb, ak}. Quite recently
another class of systems composed of bosonic modes of systems with
continuous degrees of freedom have come under intensive
investigations \cite{juls, duan, werner, plenio1, plenio2,
wolfreport, cirac, wolfpassive, oliv, illum}. There are a number
of reasons for this interest. First, bosonic modes are the
appropriate subsystems the entanglement of which must be
calculated when dealing with systems of identical boson particles
\cite{enk, gitt, zanardi, ved}. Second, states of continuous
systems are widely encountered in many branches of physics in
which entanglement plays a role, i.e. namely in quantum optical
setups, in atomic ensembles interacting with electromagnetic
fields \cite{juls}, in the motion of ions in ion traps and in low
excitations of bosonic field theories. Third, for a class of such
states, namely Gaussian states, analytical measures of
entanglement have been defined and calculated in closed form
\cite{cirac}. Finally, a large class of interesting many body
systems of this type, when written in terms of suitable
coordinates, are in fact free systems. This latter reason makes a
fuller investigation of the above questions in such systems much
easier than in spin systems. For example it has been shown
\cite{plenio1} that in the ground state of a ring of particles
coupled with each other by harmonic forces, entanglement exists
only between adjacent particles and is exactly zero if the
particles are non-adjacent. It has also been possible to study the
development of entanglement when such systems evolve in
time \cite{plenio2}.  Neither of these questions are easy to investigate in spin systems.\\
Furthermore quite interestingly the notion of entanglement
frustration has recently been introduced in \cite{wolf}. The
motivation for this notion comes from the observation that quantum
entanglement like many other local properties of a system becomes
strongly restricted when one imposes global symmetry requirements
on the state of a system of $N$ particles. For example the
Entanglement of Formation (EoF)\cite{cirac} of two modes in a
Gaussian state is generally unbounded, it can assume any value
between zero and infinity. However if these modes are part of a
three mode system having a permutation symmetry, then the amount
of their EoF becomes finite. Therefore one may ask how the EoF of
two modes in a system, may depend on the total number of modes,
the symmetry of the state and the dimension of the lattice which
represents the interaction of these modes. These questions have
been studied in \cite{wolf}, with the aim of determining the
maximum value of EoF which can exist between two modes given that
the whole state has a general symmetry. However the Hamiltonian of
\cite{wolf} is a specific hamiltonian having no tunable
parameter(i.e. frequencies), in fact it has been so defined that
its ground state allows the maximum value of EoF among all
the possible quadratic Hamiltonians with the same symmetry. \\

In this paper we want to study the effect of interaction strength
and symmetry in determining the entanglement of bosonic modes in a
class of states pertaining to the ground state of symmetric
graphs. We take a system of $N$ particles or bosonic modes coupled
quadratically with each other so that the Hamiltonian allows a
global symmetry. We then  determine the entanglement of two modes
when the system is in its ground state. The entanglement depends
on the number of particles, the strength of interaction and the
symmetry of the Hamiltonian. We will see that entanglement always
increases (almost rapidly) with the strength of interaction and
always saturates to a finite limiting value but does not exceed
the bounds found in \cite{wolf}. We also show that in some cases
symmetry has a constructive effect on entanglement. For example in
some symmetric graphs like a cube or an octahedron, there are
entanglement between next-nearest neighbors which is known to be
absent in less symmetric graphs, i.e. in rings \cite{plenio1}. In
the most simple case, where the graph consists of two vertices
connected by an edge, we also calculate the thermal entanglement
as a function of temperature and interaction strength and
determine the threshold temperature beyond which entanglement
vanishes. \\ The structure of this paper is as follows: In section
\ref{gauss} we review briefly the definition of Gaussian states
and the Entanglement of Formation (EoF) for symmetric two-mode
Gaussian states. In section \ref{potential} we introduce the
quadratic Hamiltonian and calculate the EoF of two Gaussian modes
when the whole system is in the ground state of this quadratic
hamiltonian. This hamiltonian simply describes a mass-spring
system of the kind considered in \cite{plenio1} and is different
from the one introduced in \cite{wolf}, which maximizes the EoF.
In section \ref{graphs} we present our results for various
symmetric graphs. We end the paper with a conclusion. A remark on
notation. Throughout the paper we use $E_{inf}$ for the maximum
EoF which is obtained at infinite frequencies (interaction
strength of the springs) in our hamiltonian and $E_{max}$ for the
bound discovered in \cite{wolf}.
\section{Gaussian States}\label{gauss}
In this section we discuss the rudimentary material on Gaussian
states \cite{holevo} that we need in the sequel. Reference
\cite{tut} can be consulted for a rather detailed review on the
subject of Gaussian states.\\ Let $\hat{R}:=(\hat{x}_1, \hat{x}_2,
\cdots \hat{x}_N, \hat{p}_1, \hat{p}_2, \cdots \hat{p}_N)$ be $N$
conjugate operators characterizing $N$ modes and subject to the
canonical commutation relations
$$
    [\hat{R}_k,\hat{R}_l]=i\sigma_{kl},
$$ where $
    \sigma = \left(
\begin{array}{cc}
   & I_{n} \\
  -I_{n} &
\end{array}
    \right)
$
is the $2n$ dimensional symplectic matrix and $I_n$ denotes the $n$ dimensional unit matrix.\\
A quantum state $\rho$ is called Gaussian if its characteristic
function defined as
$$
   C(\xi):= tr(e^{-i\xi_k\sigma_{kl} \hat{R}_l}\rho),
$$
is a Gaussian function of the $\xi$ variables, namely if
$$
   C(\xi):= e^{\frac{-1}{2}\xi_k\Gamma_{kl}\xi_l},
$$
where we have assumed that linear terms have been removed by
suitable unitary transformations.  The matrix $\Gamma$, called the
covariance matrix of the state, encodes all the correlations in the
form
$$
\Gamma_{kl}:= \la R_kR_l+R_lR_k\ra - 2\la R_k\ra \la R_l\ra.
$$

By symplectic transformations the covariance matrix of a two mode
symmetric Gaussian state (one which is invariant under the
interchange of the two modes) can always be put into the standard
form
\begin{equation}\label{f}
    \gamma=\left(
    \begin{array}{cccc}
      n & k_x &  &  \\
      k_x & n &  &  \\
       &  & n & k_p \\
       &  & k_p & n
    \end{array}
    \right),
\end{equation}
where $k_x\geq 0 \geq k_p$ and $k_x\geq |k_p|$. For such a state
the Entanglement of Formation (EoF) has been obtained in
\cite{cirac}. It is given by
\begin{equation}\label{g}
    EoF(\rho):= C_+\log_2 C_+-C_-\log_2 C_-,
\end{equation}
in which
\begin{equation}\label{h}
    C_{\pm} = \frac{(1\pm \Delta)^2}{4\Delta},
\end{equation}
and
\begin{equation}\label{i}
\Delta := min(1, \delta:=\sqrt{(n-k_x)(n+k_p)}).
\end{equation}
Thus a state is entangled only if $\delta\leq 1$.  In general $EoF$
is unbounded and can assume any value between $0$ for $\delta = 1 $
to infinity for $\delta \lo 0 $.

\section{The ground and thermal entanglement of a system of particles coupled by harmonic forces}\label{potential}
Consider now a system of particles with canonical variables $R=(x_1,
x_2, \cdots x_N,p_1,p_2,\cdots p_N)^T$ subject to the following
Hamiltonian
\begin{equation}\label{j}
    H = \frac{1}{2}\sum_{k=1}^N {p_k^2} + \frac{1}{2}\sum_{k=1}^N
    x_k^2 +
    \frac{1}{2}\omega^2\sum'_{k,l}(x_k-x_l)^2,
\end{equation}
where the prime over the sum indicates that only specific particles
are coupled with each other.  This Hamiltonian can be written in
matrix form
\begin{equation}\label{k}
    H  = \frac{1}{2}R^T \left(
\begin{array}{cc}
  \hat{V} & 0 \\
  0 & I
\end{array}
    \right)R,
\end{equation}
where $\hat{V}$ introduces the quadratic matrix of the potential.
The state of such a system at temperature $T=\frac{1}{\beta}$ is
given by $\rho=\frac{e^{-\beta H}}{Z}$, where $Z=tr e^{-\beta H}$ is
the partition function. It is not difficult to show that this state
is Gaussian \cite{plenio1} with covariance matrix

\begin{eqnarray}\label{l}
    \Gamma  &=&\Gamma_x\oplus \Gamma_p,\cr
    \Gamma_x &=& W^{-1}\coth \frac{\beta W}{2},\cr
    \Gamma_p &=& W\coth \frac{\beta W}{2},
\end{eqnarray}

where $W:=V^{\frac{1}{2}}$. At zero temperature when only the ground
state is populated the covariance matrix tends to

\begin{equation}\label{o}
    \Gamma (T=0) =W^{-1}\oplus W.
    \end{equation}
Once the covariance matrix of the $N$ modes is obtained as above,
the covariance matrix pertaining to any two modes, say the modes $1$
and $2$ is given by the sub-matrix obtained from $\Gamma $ by
deleting the rows and columns corresponding to all the other modes.
Furthermore if this Gaussian state is symmetric, i.e. if it is
invariant under the interchange of the two modes, then its
covariance matrix will have the following (not-yet standard) form
\begin{equation}\label{p}
    \gamma=\left(
\begin{array}{cccc}
  n_x & m_x  &  &  \\
  m_x & n_x &  &  \\
  &  & n_p &m_p  \\
   &  &m_p  & n_p
\end{array}
    \right),
\end{equation}
where
\begin{eqnarray}\label{q}
    &&n_x:=(\Gamma_x)_{11}=(\Gamma_x)_{22}\cr
    &&n_p:=(\Gamma_p)_{11}=(\Gamma_p)_{22}\cr
    &&m_x:=(\Gamma_x)_{12}=(\Gamma_x)_{21}\cr
    &&m_p:=(\Gamma_p)_{12}=(\Gamma_p)_{21}.
\end{eqnarray}
By the canonical transformation
\begin{equation}\label{r}
x_{1,2}\lo \alpha x_{1,2},\h p_{1,2}\lo \frac{1}{\alpha} p_{1,2},
\end{equation}
where $\alpha = (\frac{n_p}{n_x})^{\frac{1}{4}}$, the covariance
matrix (\ref{p}) takes the standard form

\begin{equation}\label{s}
    \gamma=\left(
\begin{array}{cccc}
  n & k_x  &  &  \\
  k_x & n &  &  \\
  &  & n &k_p  \\
   &  &k_p  & n
\end{array}
    \right),
\end{equation}
where

\begin{eqnarray}\label{t}
    n&:=&\sqrt{n_xn_p},\cr
    k_x&:=& m_x\sqrt{\frac{n_p}{n_x}},\cr
    k_p&:=& m_p\sqrt{\frac{n_x}{n_p}}.
\end{eqnarray}
From this standard form and (\ref{i}) one can now determine the
EoF of the
two modes.\\
Inserting (\ref{t}) in (\ref{i}) we find:
\begin{eqnarray}\label{tt}
    \delta^2 &=& (n-k_x)(n+k_p)\cr &=&
    (\sqrt{n_xn_p}-m_x\sqrt{\frac{n_p}{n_x}})(\sqrt{n_xn_p}+m_p\sqrt{\frac{n_x}{n_p}})\cr
    &=&  (n_x-m_x)(n_p+
    m_p),
    \end{eqnarray}
which in view of (\ref{q}) leads to
\begin{equation}\label{u}
   \delta=\sqrt{\left((\Gamma_x)_{11}-(\Gamma_x)_{12}\right)\left((\Gamma_p)_{11}+
    (\Gamma_p)_{12}\right)}.
\end{equation}
The matrix elements of $\Gamma_x$ and $\Gamma_p$ are obtained  by
diagonalization of $V$. At zero temperature (i.e. for the ground
state) we will have
\begin{eqnarray}\label{v}
    \delta &=&
    \sqrt{\left((W^{-1})_{11}-(W^{-1})_{12}\right)\left(W_{11}+W_{12}\right)}.
\end{eqnarray}
In the following section we will apply these results to evaluate the
EoF of states for modes pertaining to various symmetric lattices.

\section{Entanglement in symmetric graphs}\label{graphs} Let $G$ be a symmetric graph, having $N$ vertices. The
adjacency matrix of the graph is denoted by $A$, where by
definition $A_{ij}=1 $ if the the vertices $i$ and $j$ are linked
by an edge and $A_{ij}=0$ otherwise. In a concrete situation we
can think of the particles of unit mass located at the vertices of
the graph experiencing a global harmonic potential and coupled
through springs of strength $k=\omega^2$ with each other. If the
number of nearest neighbors of a vertex is denoted by $z$, then
the potential matrix $\hat{V}$ takes the following form
 \be\label{w}
 \hat{V} = (1+z\omega^2)I-\omega^2 A.
 \ee
In the following we will consider various symmetric graphs and
obtain in each case the EoF of the two modes corresponding to
various pair of vertices. In almost all cases calculations show
that entanglement exist only between nearest neighbor sites. The
only exceptions are the octahedron and the cube where the
next-nearest and next-next-nearest neighbors are also entangled
with each other. In the simplest case, where the lattice consists
of only two vertices connected by an edge, we also determine the
threshold temperature above which entanglement disappears. This
case may model the entanglement between the vibrational modes of
two atoms in a molecule or two ions in an ion trap which is
destroyed by thermal fluctuations.

\subsection{The simplest two-vertex graph}
\begin{figure}[t]
 \centering
   \includegraphics[width=8cm,height=8cm,angle=0]{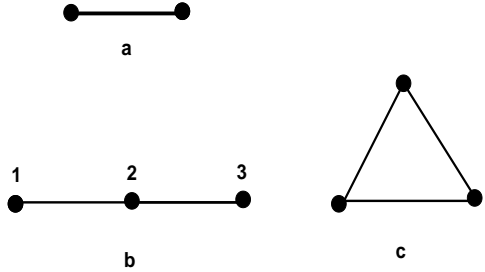}
   \caption{The simplest symmetric graphs. The graph in (b) is only symmetric with respect to the interchange
   of the vertices
   1 and 3. The EoF in graph (a) is unbounded, that of graph (c) is bounded but much higher than
    that of vertices 1 and 3 in graph (b).}
   \label{new3}
\end{figure}

The simplest symmetric graph consists of two vertices connected by
an edge, figure (\ref{new3}-a).
The potential for this graph is
$V=\frac{1}{2}(x_1^2+x_2^2+\omega^2(x_1-x_2)^2)$. It represents the
harmonic interaction of two atoms with frequency $\omega$ each
trapped inside a harmonic well of unit frequency. The potential
matrix is
\begin{equation}\label{x}
    \hat{V} = \left(
\begin{array}{cc}
  1+\omega^2 & -\omega^2 \\
  -\omega^2 & 1+\omega^2
\end{array}
    \right),
\end{equation}
 with eigenvalues and eigenvectors
\begin{eqnarray}\label{y}
    \omega_1^2 &:=& 1, \h \ \ \ \ \ \ \ \ |e_1\ra = \frac{1}{\sqrt{2}}\left(\begin{array}{c}
                                       1 \\
                                       1
                                     \end{array}\right),\cr
\omega_2^2&:=& 1+2\omega^2, \h |e_2\ra =
\frac{1}{\sqrt{2}}\left(\begin{array}{c}
                                       1 \\
                                       -1
                                     \end{array}\right).
\end{eqnarray}

\begin{figure}[t]
\centering
   \includegraphics[width=8cm,height=8cm,angle=0]{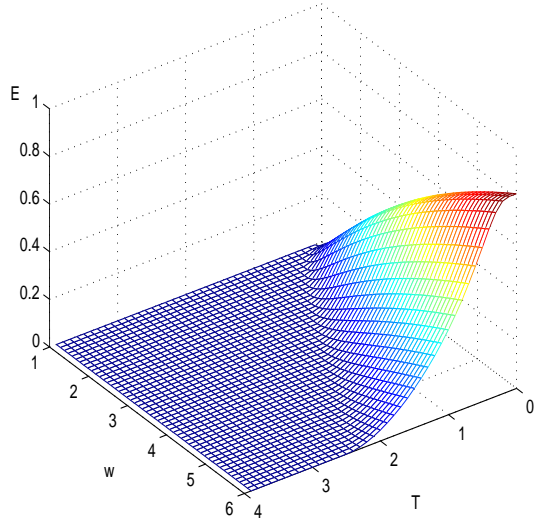}
   \caption{Thermal entanglement of the two modes (particles) (in units of ebits) associated to the graph in figure
   (\ref{new3}-a),
as a function of temperature and frequency. $E$ is dimensionless
and we are working in units in which $\omega$ and $T$ are also
dimensionless.}
   \label{thermal2}
\end{figure}

This leads to the following values for the relevant matrix elements:
\begin{eqnarray}\label{a1}
    ({\Gamma_x})_{11} &=& \frac{1}{2} \left(
\coth \frac{\beta}{2} + \frac{1}{\omega_2}\coth
\frac{\beta\omega_2}{2}\right),\ \ \ \ ({\Gamma_x})_{12} =
\frac{1}{2} \left( \coth \frac{\beta}{2} - \frac{1}{\omega_2}\coth
\frac{\beta\omega_2}{2}\right),\cr
  ({\Gamma_p})_{11} &=& \frac{1}{2} \left(
\coth \frac{\beta}{2} + \omega_2\coth
\frac{\beta\omega_2}{2}\right),\ \ \ \ ({\Gamma_p})_{12} =
\frac{1}{2} \left( \coth \frac{\beta}{2} - \omega_2\coth
\frac{\beta\omega_2}{2}\right).
   \end{eqnarray}
Inserting these values in (\ref{u}) we find the value of $\delta$,

\begin{equation}\label{a1extra}
    \delta = \sqrt{\omega_2^{-1}{\coth
\frac{\beta}{2}\coth \frac{\beta\omega_2}{2}}}
\end{equation}
The equation $\delta = 1 $ determines the threshold temperature.
Inserting this value of delta in (\ref{g}) and (\ref{h})  will
determine the EoF as a function of the frequency $\omega$ and the
temperature, shown in figure (\ref{thermal2}). The maximum
entanglement obtains at zero temperature and is unbounded, it
behaves as $E = -0.56+\log_2(w)+O(\omega^{-2})$ for
large frequencies. The threshold temperature increases almost linearly with frequency. \\

\subsection{The simplest three-vertex symmetric graphs}
We now consider the graph shown in figure (\ref{new3}-b).
 This is one of the
two symmetric graphs with three vertices. The other graph which is
a triangle will be treated as a special case of $N$-vertices mean
field clusters (for $N=3$). Note that this graph is symmetric only
under the interchange of vertices $1$ and $3$. We will determine
the EoF of these two modes. Our motivation for studying this graph
is to show that entanglement can exist between next-nearest
neighbors in open chains with an odd number of vertices, a
property which we have seen also for higher than $3$ vertices in
our calculations. Here the potential is

\begin{equation}\label{a2}
    V=\frac{1}{2}(x_1^2+x_2^2+x_3^2)+\frac{\omega^2}{2}((x_1-x_2)^2+(x_2-x_3)^2),
\end{equation}
with the potential matrix
\begin{equation}\label{a3}
    \hat{V} = \left(
\begin{array}{ccc}
  1+\omega^2 & -\omega^2 & 0 \\
  -\omega^2 & 1+2\omega^2 & -\omega^2 \\
   0& -\omega^2 & 1+\omega^2
\end{array}\right).
\end{equation}
The eigenvalues and eigenvectors are
\begin{eqnarray}\label{a4}
  \omega_1^2&=&1, \h  \ \ \ \ \ \ \ \ |e_1\ra = \frac{1}{\sqrt{3}}(1,1,1)^T, \cr
  \omega_2^2&=&1+\omega^2, \h \ |e_2\ra = \frac{1}{\sqrt{2}}(1,0,-1)^T, \cr
  \omega_3^2&=& 1 + 3\omega^2, \h |e_3\ra =
  \frac{1}{\sqrt{6}}(1,-2,1)^T.
\end{eqnarray}
In this case we are only interested in the ground state
entanglement. So only the matrix elements of $W$ and $W^{-1}$ need
be calculated. One finds after straightforward calculations along
the lines in the previous subsection the following value for
$\delta $,
\begin{equation}\label{a5}
    \delta=\sqrt{\frac{2+\sqrt{1+3\omega^2}}{3\sqrt{1+\omega^2}}}.
\end{equation}
The value of this parameter is always less than $1$, and hence
there is entanglement between the next-nearest neighbors at all
frequencies. The interesting point is however that this
entanglement is bounded. It obtains at very large frequencies
where $\delta \lo 3^{-\frac{1}{4}}$, leading to $E_{inf} = 13.62$
in units of $10^{-2}$ ebits.

\subsection{Mean field clusters}
\begin{figure}[t]
\centering
   \includegraphics[width=8cm,height=8cm,angle=-90]{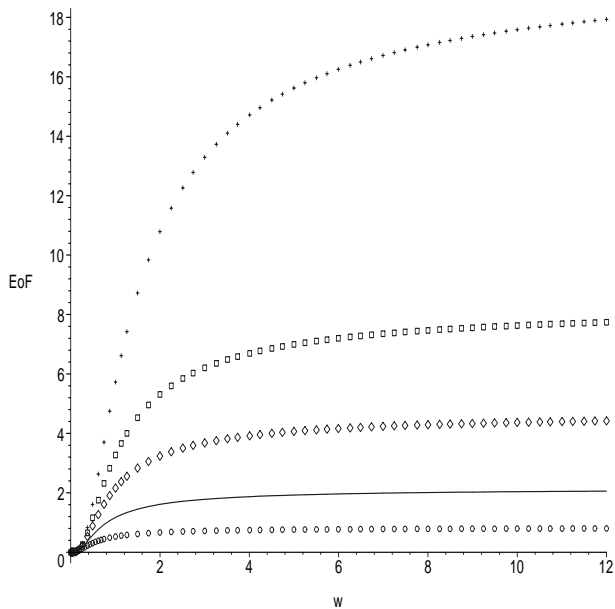}
   \caption{The EoF (in units of $10^{-2}$ ebits) of two vertices for mean field graphs with number
   of vertices from top to bottom $N=4,6,8,12$ and $20$.} \label{meanfield4681220}
\end{figure}

Consider a mean field cluster of $N$ vertices in which every
vertex is connected to $N-1$ other vertices. The potential matrix
for this graph is given by

\be\label{a6} \hat{V} = (1+N\omega^2)I - \omega^2\hat{E}, \ee
where $E$ is the matrix  all of whose entries are equal to 1, $
    E_{ij}=1 \ \ \ \forall\ \ \  i,\ \ $  and $\  j $. Using the property $E^2 = NE$,  we find
\begin{eqnarray}\label{a7}
   W &=& \sqrt{1+N\omega^2}I + \frac{1-\sqrt{1+N\omega^2}}{N}E \cr
   W^{-1}&=& \frac{1}{\sqrt{1+N\omega^2}}(I +
   \frac{\sqrt{1+N\omega^2}-1}{N}E).
\end{eqnarray}
Using these matrices and equation (\ref{v}) one finds the value of
$\delta$ at zero temperature
\begin{equation}\label{a9}
    \delta_N =
    \sqrt{\frac{2+(N-2)\sqrt{1+N\omega^2}}{N\sqrt{1+N\omega^2}}}
\end{equation}
Inserting this into equations (\ref{g}) and (\ref{h}) gives the
entanglement as a
function of $N$ and frequency. \\
Let us first consider the special case of $3$ vertices, whose
graph is a triangle, shown in figure (\ref{new3}-c) and compare it
with
result for the three-vertex graph shown in figure ( \ref{new3}-b).\\
For $N=3$, we have $\delta_3
=\sqrt{\frac{2+\sqrt{1+3\omega^2}}{3\sqrt{1+3\omega^2}}}$. The Eof
saturates at $E_{inf}=40.08$ (in units of $10^{-2}$ ebits) for
infinite
frequencies. \\
Returning to (\ref{a9}) we can find the EoF in mean field clusters
as a function of frequency for fixed number of vertices. The
maximum EoF always occurs for infinitely large frequencies where
$\delta$ tends to $\sqrt{\frac{N-2}{N}}$. Figure
(\ref{meanfield4681220} ) shows the EoF as a function of $w$ for
some mean field clusters with $N=4,6,8,12$ and $20$ vertices.
These are the number of vertices of regular polyhedra, namely
tetrahedron, octahedron, cube, dodecahedron and isocahedron  respectively, which will be treated in the nest subsection. \\

\begin{figure}[t]
\centering
   \includegraphics[width=8cm,height=8cm,angle=-90]{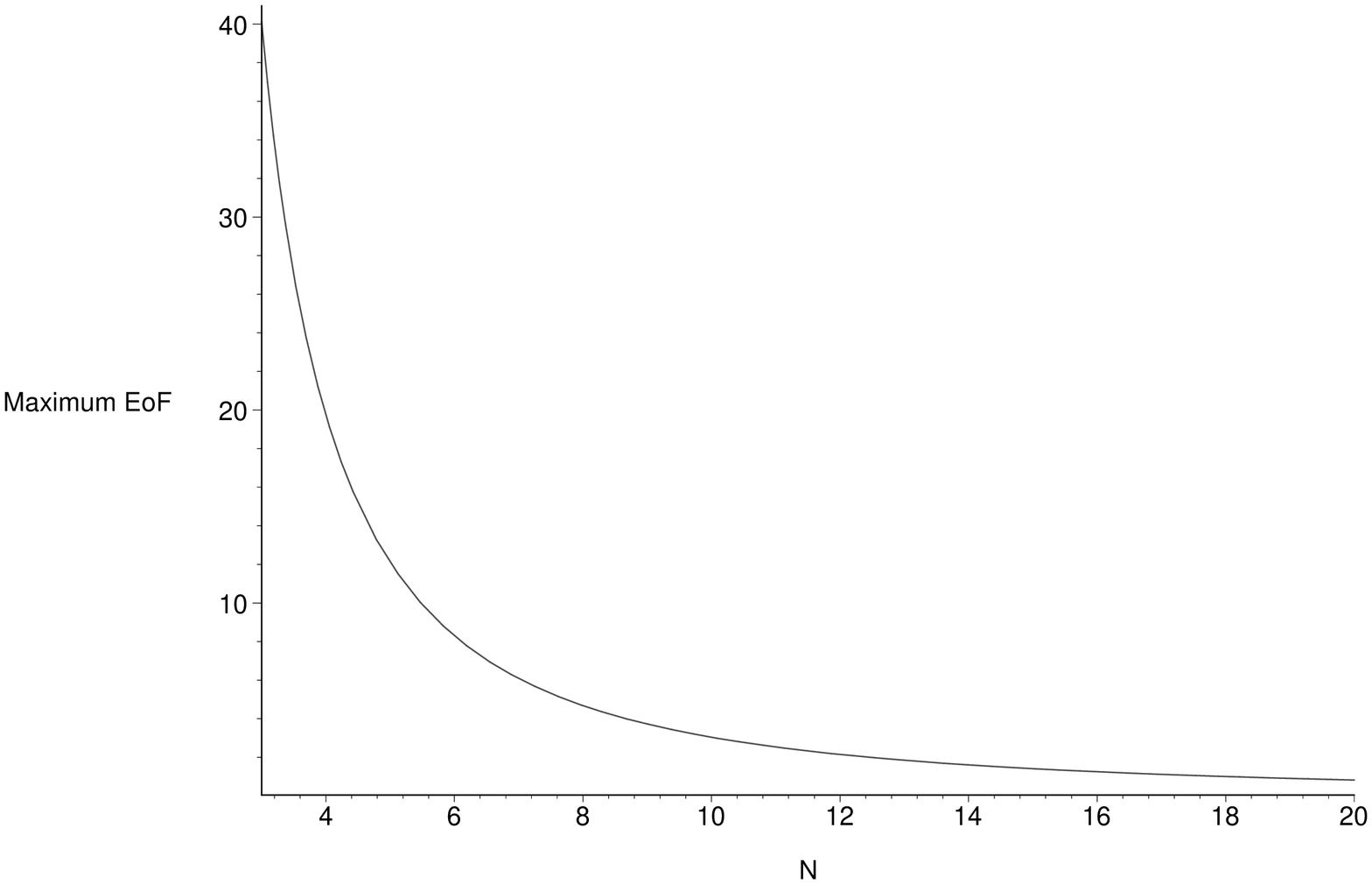}
   \caption{The maximum EoF (in units of $10^{-2}$ ebits) for mean field clusters as a function of the number of vertices.}
   \label{meanfieldEmax}
\end{figure}

Finally figure(\ref{meanfieldEmax}) shows how the maximum EoF
decreases with increasing the number of vertices in mean field
clusters. For large values of $N$ the EoF vanishes as $E\approx
\frac{1.72\log N}{2N^2}$.
\subsection{Regular Polyhedra}
\begin{figure}[t]
\centering
   \includegraphics[width=8cm,height=8cm,angle=-90]{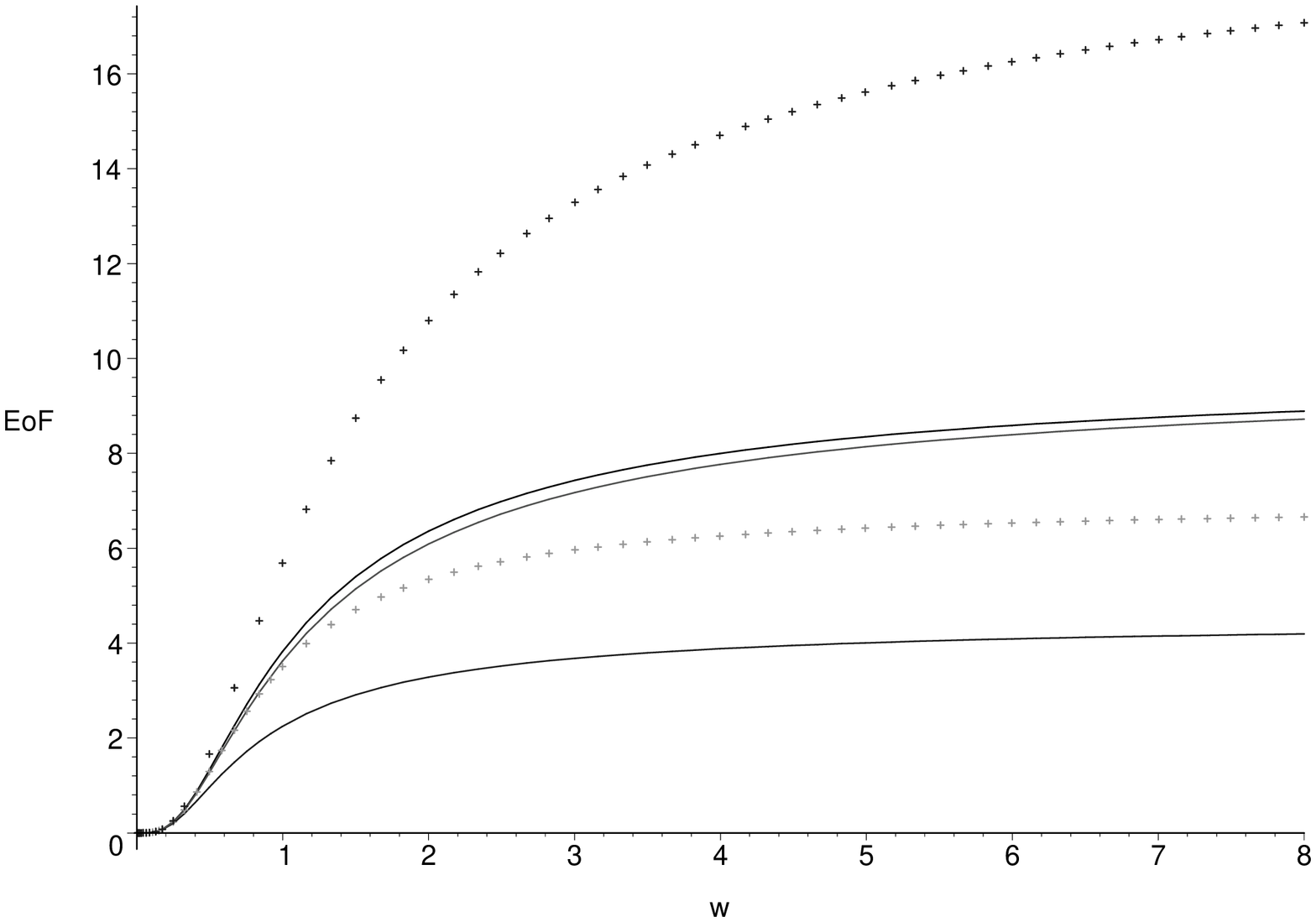}
   \caption{The EoF of adjacent vertices (in units $10^{-2}$ ebits) in regular polyhedra.  The curves correspond from top to bottom to
   Tetrahedron, Cube, Octahedron, Dodecahedron and Isocahedron.} \label{polyhedra2}
\end{figure}

\begin{figure}[t]
\centering
   \includegraphics[width=8cm,height=8cm,angle=-90]{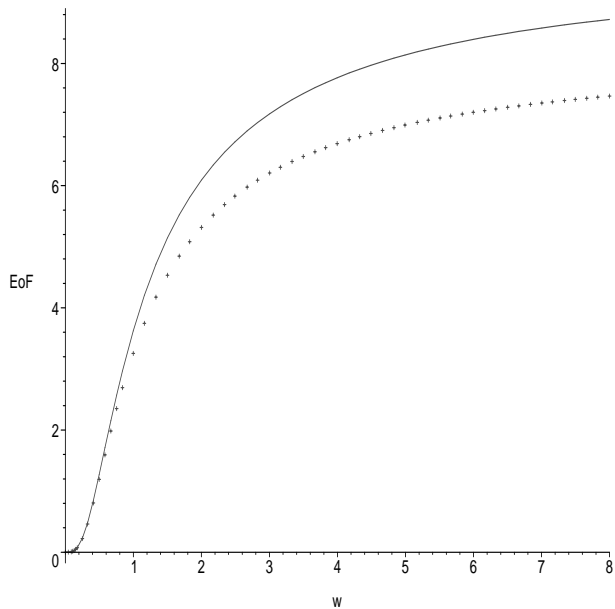}
   \caption{The EoF of adjacent vertices (in units $10^{-2}$ ebits) as a function of $w$ for the regular octahedron
   (line)
   and a six-vertex mean field cluster (dots). } \label{octmean6}
\end{figure}

In this section we consider the regular polyhedra, the tetrahedron
($N=4$), the cube ($N=8$), the octahedron ($N=6$), the
dodecahedron ($N=20$) and the isocahedron ($N=12$). The
tetrahedron is a mean field cluster with $N=4$, so we can use the
results of the mean field cluster for this graph. For the other
cases we follow the same procedure as above, i.e. determine the
adjacency matrix which in turn determines the potential matrix
from (\ref{w}). The square root $W$ and the inverse square root
$W^{-1} $ of $\hat{V}$ are calculated by diagonalization of
$\hat{V}$. We can then calculate the necessary elements of $W$ and
$W^{-1}$ in order to determine the EoF of pair of vertices in
terms of the strength of the interaction $\omega$. It turns out
that the $EoF$ is always an increasing function of $\omega$ and
saturates for infinite frequencies. The maximum EoF attains its
value for the nearest neighbors. Depending on the polyhedron, the
EoF may or may not exist between next nearest neighbors.
Furthermore compared with mean field clusters the EoF is always
higher than that of mean field clusters, the difference is larger
for higher frequencies. Figure (\ref{polyhedra2}) shows the EoF of
the regular polyhedra in terms of the strength of interaction.
Figure (\ref{octmean6}) compares the EoF of the octahedron with
that of a mean field cluster with $6$ vertices. Finally table (1)
compares the maximum EoF obtained in regular polyhedra with those
of the mean field clusters of the same size (the same number of
vertices) and the bounds found in \cite{wolf}.
\vskip 1cm
\vskip 1cm \begin{center}
\begin{tabular}{|c|c|c|c|c|}
  \hline
  {\rm Polyhedron} & N&{\rm Mean field Cluster} &$E_{{\rm inf}}$& $E_{max}$ \\
  \hline
  {\rm Tetrahedron} &4 &19.74 &19.74& 19.74 \\
  {\rm Cube} &8&8.30  &9.80&19.74 \\
  {\rm Octahedron} &6&4.68& 9.74& 10.75 \\
  {\rm Dodecahedron} &20&  2.15&7.00& 11.12 \\
  {\rm Isocahedron} &12 & 0.83 &4.51&5.37 \\
  \hline
\end{tabular}
\end{center}
\begin{center}
Table 1-The maximum entanglement ($E_{inf}$) in regular polyhedra
and mean field graphs of the same size, compared with the bounds
found in \cite{wolf}. $N$ is the number of vertices of the
polyhedron and the mean field graph.
\end{center}

\begin{center}

\begin{tabular}{|c|c|c|c|c|c|}
  \hline
  {\rm Polyhedron} & $E^1_{{\rm inf}}$& $E^2_{{\rm inf}}$ & $E^3_{\rm{inf}}$&$E^4_{\rm{inf}}$& $E^5_{\rm{inf}}$ \\
  \hline
  {\rm Tetrahedron} &19.74 &- & - &-&-\\
  {\rm Cube} &9.80&1.08 &0.24 &-&-\\
  {\rm Octahedron} &9.74&2.58 & - &-&-\\
  {\rm Dodecahedron} &7.00& 0 & 0 &0&0\\
  {\rm Isocahedron} &4.51 &0 & 0 &0&0\\
  \hline
\end{tabular}
\end{center}
\begin{center}
Table 2-The EoF between different vertices in regular polyhedra.
The superscript indicates the minimum distance between the
vertices in terms of the number of edges. The $-$ sign indicates
that such distances do not occur in the corresponding regular
polyhedron.
\end{center}
\subsection{Rings and rectangular lattices}
\begin{figure}[t]
\centering
   \includegraphics[width=8cm,height=8cm,angle=-90]{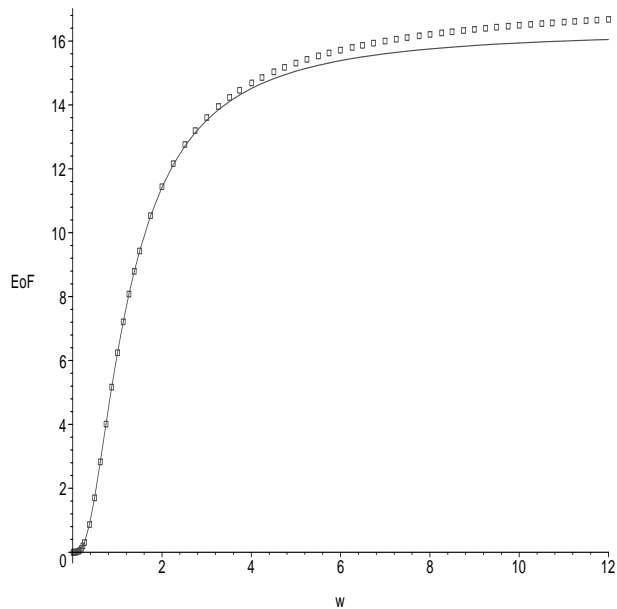}
   \caption{The EoF of two adjacent sites for rings with different number of vertices almost coincide for low frequencies, $N=11$ (points)
   $N=201$ (line).} \label{rings-low}
\end{figure}

\begin{figure}[t]
\centering
   \includegraphics[width=8cm,height=8cm,angle=-90]{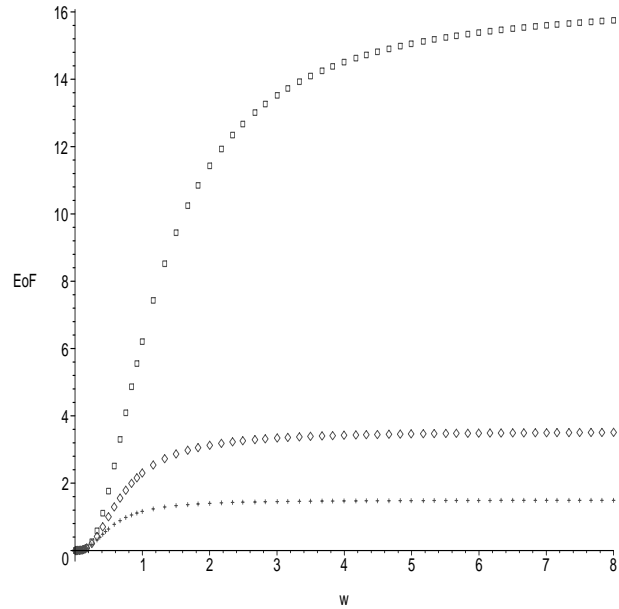}
   \caption{The EoF for infinite dimensional lattices as a function of frequency, from top to bottom $d=1,d=2$ and $d=3$.}
   \label{ringtoruscube}
\end{figure}

 For the $d$-dimensional cubic lattice, of $N^d$
vertices, the potential matrix is
\begin{equation}\label{a10}
\hat{V} = (1+2d\omega^2)I - \omega^2
\sum_{k=1}^d(\hat{T_k}+\hat{T_k}^{-1}),
\end{equation}
where $T_k$ is the cyclic unit shift operator in the $k-$ direction.
The eigenvalues and eigenvectors are as follows:
\begin{equation}\label{a11}
\omega_{s}^2={1+4\omega^2 \sum_{k=1}^d\sin^2\frac{\pi s_k}{N}},\h
u_{\bf s}({\bf n}):=\frac{1}{\sqrt{N^d}} e^{\frac{2\pi i {\bf s\cdot
n}}{N}}.
\end{equation}
From these values one can easily obtain the relevant matrix elements
of $W$ and $W^{-1}$. For example for the one dimensional lattice we
will have
\begin{equation}\label{a12}
 {W^{\pm 1}}_{k,l} = \sum_{s=0}^{N-1}({1+4\omega^2 \sin^2 \frac{\pi
  s}{N}})^{\frac{\pm 1}{2}}e^{\frac{2\pi i s(k-l)}{N}}.
  \end{equation}
 As always the EoF increases with the strength of interaction. It
turns out that EoF exists only between nearest neighbor sites.
Figure (\ref{rings-low}) compares the EoF of rings with two very
different size as a function of frequency. The interesting point
is that, the difference for the number of vertices only shows
itself when the frequency is high
enough.\\
Figure (\ref{ringtoruscube}) compares the EoF of $d-$ dimensional
lattices for $d=1,2,3$ when $N\lo \infty$.
 For the infinite lattices, the sums in (\ref{a12})
turn into integrals and we will have for two adjacent vertices say
$1$ and $2$ of one dimensional lattice,
\begin{eqnarray}\label{a13}
    W^{\pm 1}_{11}&=&\frac{1}{\pi}\int_{0}^{\pi} dx ({1+4\omega^2 \sin^2
    x})^{\pm\frac{1}{2}}\cr
    W^{\pm 1}_{12}&=&\frac{1}{\pi}\int_{0}^{\pi} dx ({1+4\omega^2 \sin^2
    x})^{\pm\frac{1}{2}}\cos{2x},
\end{eqnarray}
and for the two dimensional lattice
\begin{eqnarray}\label{a14}
    W^{\pm 1}_{11}&=&\frac{1}{\pi^2}\int_{0}^{\pi}dx \int_{0}^{\pi}dy [{1+4\omega^2 (\sin^2
    x + \sin^2 y})]^{\pm\frac{1}{2}}\cr
    W^{\pm 1}_{12}&=&\frac{1}{\pi^2}\int_{0}^{\pi}dx\int_{0}^{\pi}dy [{1+4\omega^2 (\sin^2
    x+\sin^2 y})]^{\pm\frac{1}{2}}\cos{2x}.
\end{eqnarray}
Table (3) compares the saturation values of EoF achieved in these
cases with the maximum achievable EoF's derived in \cite{wolf}.
\begin{center}
\begin{tabular}{|c|c|c|}
  \hline
  {\rm d} & $E_{{\rm inf}}$& $E_{{\rm max}}$ \\
  \hline
  1 &16.34 &30 \\
  2&3.54&6.31  \\
  3 &1.50&2.62 \\
  \hline
\end{tabular}
\end{center}
\begin{center}
Table 3-The maximum EoF between adjacent vertices in d-dimensional
rectangular lattices compared with the bounds found in
\cite{wolf}.
\end{center}

\section{Discussion}
We have investigated the entanglement properties of the ground
state of a quadratic Hamiltonian governing the harmonic
interaction of bosonic modes or particles located on the vertices
of symmetric graphs i.e. 1,2, and 3 dimensional rectangular
lattices, mean field clusters and platonic solids. The
entanglement has been calculated as a function of the interaction
strength. In each case the EoF is an increasing function of the
interaction strength and it saturates to
 a finite value dictated by the type of the graph. Our results confirm those of \cite{wolf} in that the maximum values that
 we obtain are always less than the bounds found in that paper which correspond to a particular quadratic hamiltonian proved to
 allow for maximum achievable EoF between any two modes in the class of all quadratic
 hamiltonians. Some peripheral results may be interesting. First, in some symmetric graphs ( the cube, the octahedron and open chains with an odd number of vertices) there are entanglement between
 non-adjacent vertices, a property which is extraordinary for
 entanglement, since for rings it has been shown \cite{plenio1} that
 entanglement can not extend beyond the nearest neighbors. Second,
 in the case of two atoms vibrating in a molecule or ions
 vibrating in an ion trap, corresponding to the simple graph
 (\ref{new3}-a) the threshold temperature above
 which entanglement is destroyed, has been determined.

\section{Acknowledgement}
We would like to thank A. Bayat, I. Marvian, D. Lashkari, N. Majd,
L. Memarzadeh, and A. Sheikhan for valuable discussions.
{}
\end{document}